\documentclass[aps,prb,superscriptaddress,twocolumn,floatfix]{revtex4}
\usepackage{graphics}
\usepackage{epsfig}

\newcommand{\kpar}{{\bf k}_{\parallel}}
\newcommand{\kparrm}{k_{\parallel}}

\begin{document}

\title{Effect of interface bonding on spin-dependent tunneling from the oxidized Co surface}

\author{K. D. Belashchenko}
\author{E. Y. Tsymbal}
\affiliation{Department of Physics and Astronomy and Center for
Materials Research and Analysis, University of Nebraska, Lincoln,
Nebraska 68588}

\author{M. van Schilfgaarde}
\affiliation{Department of Chemical and Materials Engineering, Arizona State University, Tempe, Arizona 85287}

\author{D. A. Stewart}
\affiliation{Sandia National Laboratories, Livermore, California 94551}

\author{I. I. Oleynik}
\affiliation{Department of Physics, University of South Florida, Tampa, Florida 33620}

\author{S. S. Jaswal}
\affiliation{Department of Physics and Astronomy and Center for
Materials Research and Analysis, University of Nebraska, Lincoln,
Nebraska 68588}

\begin{abstract}
We demonstrate that the factorization of the tunneling
transmission into the product of two surface transmission
functions and a vacuum decay factor allows one to generalize
Julli\`ere's formula and explain the meaning of the ``tunneling
density of states'' in some limiting cases. Using this
factorization we calculate spin-dependent tunneling from clean and
oxidized fcc Co surfaces through vacuum into Al using the
principal-layer Green's function approach. We demonstrate that a
monolayer of oxygen on the Co (111) surface creates a spin-filter
effect due to the Co-O bonding which produces an additional
tunneling barrier in the minority-spin channel. This changes the
minority-spin dominated conductance for the clean Co surface into
a majority spin dominated conductance for the oxidized Co surface.
%reverses the sign of the SP from negative for the clean
%Co surface to positive for the oxidized Co.
\end{abstract}

%\begin{keyword}
%spin-dependent tunneling \sep principal-layer Green's function approach \sep magnetic tunnel junctions
%\end{keyword}

\maketitle

\section{Introduction}

Spin-dependent tunneling (SDT) in magnetic tunnel junctions (MTJs)
is a dynamically developing area of research that attracted a lot
of attention due to promising applications in non-volatile random
access memories and next-generation magnetic field sensors (for a
recent review of SDT see Ref.~\onlinecite{TML}). The experimental
efforts have succeeded in achieving large reproducible tunneling
magnetoresistance (TMR) in MTJs~\cite{Moodera} but also raised
fundamental questions regarding the nature of SDT. One such
question is the role of the ferromagnet/insulator interfaces in
controlling the spin polarization (SP) of the tunneling
conductance. The SP is usually defined as
$P=(G_\uparrow-G_\downarrow)/(G_\uparrow+G_\downarrow)$, where
$G_\sigma=(e^2/h)\sum_{\kpar} T_\sigma(\kpar)$ is the conductance
for spin channel $\sigma$, $T_\sigma$ is the transmission
function, and $\kpar$ is the lateral wave vector.

Commonly the expected spin dependence of the tunneling current is
deduced by considering the symmetry of the Bloch states in the
bulk ferromagnetic electrodes and the complex band structure of
the insulator. \cite{Maclaren,Mavro} By identifying those bands in
the electrodes that are allowed by symmetry to couple efficiently
to the evanescent states decaying most slowly in the barrier one
can predict the SP of the conductance. However, this approach has
two deficiencies. First, it assumes that the barrier is
sufficiently thick so that only a small focused region of the
surface Brillouin zone (SBZ) contributes to the tunneling current.
For realistic MTJs with a barrier thickness of about 1~nm this
assumption is usually unjustified. Second, symmetry considerations
alone applied to bulk materials are not always sufficient to
predict the SP. It is critical to take into account the electronic
structure of the ferromagnet/barrier interfaces which, as it was
shown both experimentally \cite{Leclair} and theoretically,
\cite{UL} controls SDT.

An important mechanism by which the interfaces affect the SP of
the tunneling current is the bonding between the ferromagnetic
electrodes and the insulator. \cite{TP97} This effect was put
forward to explain positive and negative values of TMR depending
on the applied voltage in MTJs with Ta$_2$O$_5$ and
Ta$_2$O$_5$/Al$_2$O$_3$ barriers \cite{Sharma} and to elucidate
the inversion of the SP in
Co/SrTiO$_3$/La$_{0.67}$Sr$_{0.33}$MnO$_3$ MTJs. \cite{Deteresa}
So far there are no theoretical studies explaining the microscopic
origin of this phenomenon.

In this paper we report the results of first-principles
calculations of SDT from clean and oxidized Co surfaces through
vacuum into Al and demonstrate the crucial role of the bonding
between Co and O atoms. This system was chosen for investigation
because the MTJs based on alumina have predominantly O-terminated
Co/Al$_2$O$_3$ interfaces. \cite{oxidized} By replacing alumina by
vacuum we can ignore the complexity of the atomic structure of the
amorphous alumina and focus on effects of surface oxidation.
Moreover, this system can be directly studied using spin-polarized
STM. \onlinecite{SP-STM}
%By representing the transmission through a sufficiently thick vacuum barrier as a product of two surface
%transmission functions and a vacuum decay factor,
%We calculate the SP of the tunneling conductance from
%clean and oxidized Co electrodes to an Al electrode in the spirit of Meservey-Tedrow experiments \cite{MT}.

We show that a monolayer of oxygen on the Co surface creates a
spin-filter effect due to the Co-O bonding by producing an
additional tunneling barrier in the minority-spin channel. This
reverses the sign of the SP from negative for the clean Co surface
to positive for the oxidized Co surface, thus revealing the
crucial role of interface bonding in SDT.

\section{Surface transmission functions}

We approach the tunneling problem in the spirit of perturbation
theory. \cite{Harrison} We consider the system consisting of
``left'' and ``right'' leads separated by a relatively thick
barrier and assume that two-dimensional translational periodicity
in transverse directions is preserved for both electrodes
including their surfaces (although it may be different for each
subsystem). Every Bloch wave with a lateral quasi-wave vector
$\kpar^L$ coming from the left lead has a decay tail in the vacuum
composed of the waves with transverse wave vectors $\kpar^L+{\bf
G}_i$ where ${\bf G}_i$ are the reciprocal lattice vectors of the
SBZ of the left lead. \cite{Mazin} At sufficient distances from
the surface (typically just a few lattice parameters for low-index
surfaces) all waves with ${\bf G}_i\neq0$ vanish and may be
neglected in the Fourier expansion of the Bloch wave in the
vacuum. On the other hand, the quasi-wave vector $\kpar^R$ is the
good quantum number for the right lead. This means that $\kpar$ is
conserved across the entire system even if there is no common
in-plane periodicity and $\kpar^L$, $\kpar^R$ are defined with
respect to different SBZs. In this case, each tunneling eigenstate
is almost identical to an evanescent plane wave in the central
region of the barrier.

\begin{figure}
\begin{center}
\epsfig{file=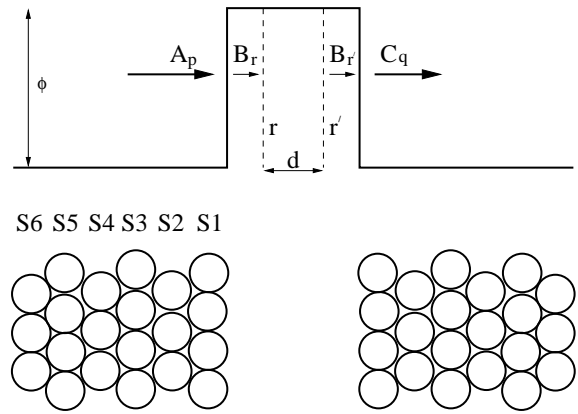,width=.42\textwidth}
\end{center}
\caption{Geometry of a tunnel junction with a vacuum barrier. The
top graph schematically shows the potential barrier for electrons
at the Fermi level. Dashed lines show the positions of the
reference planes assigned to the electrodes for the calculation of
the surface transmission functions. Each scattering state is
described by four amplitudes: $A_p$ of the incoming Bloch wave;
$B_r$, $B_r'$ of the surviving evanescent wave at the reference
planes assigned to the left and right electrodes; and $C_q$ of the
transmitted Bloch wave in the right electrode (there may be
several transmitted waves). The bottom scheme depicts atomic
layers in the electrodes and their labelling used in the text.}
\label{reference}
\end{figure}

For a given $\kpar$ the transmission function is the sum of the
transmission coefficients for all tunneling eigenstates
corresponding to all incoming Bloch waves with this $\kpar$ in the
left electrode. At the same time, each transmission coefficient
for a given eigenstate contains a sum over outgoing states in the
right electrode with the same $\kpar$. Let us choose a reference
plane in the vacuum region at a sufficient distance from the
surface of an electrode, so that the eigenstates for all $\kpar$
are already indistinguishable from the barrier eigenstates at this
plane (see Fig. \ref{reference}). For each tunneling eigenstate
the amplitude of the barrier eigenstate between the reference
planes is the only parameter coupling the left and right
electrodes. Then, the S-matrix element coupling the states in the
two electrodes may be written as
\begin{equation}\label{Smatrix}
S^\sigma_{pq}=S^\sigma_{pr}S_{rr^\prime}S^\sigma_{r^\prime q}
\end{equation}
where $p$ is the incoming and $q$ the outgoing Bloch states in the
left and right leads respectively, $r$ and $r^\prime$ denote the
same vacuum eigenstate at the left and right reference planes. We
omitted the dependence on $\kpar$ for all the S-matrices in Eq.
(\ref{Smatrix}) for brevity. The vacuum S-matrix $S_{rr^\prime}$
simply describes the exponential decay of the wave function in the
vacuum. Note that no summation is implied in (\ref{Smatrix}),
because the state $r$ is uniquely defined by $\kpar$. The simple
product of S-matrices in Eq. (\ref{Smatrix}) without any multiple
scattering terms is a consequence of our assumption that the
barrier is sufficiently thick. Thus, we see that the transmission
function $T(\kpar)$ of the MTJ is factorized:
\begin{equation}
T_{\sigma}(\kpar)=t^{\sigma}_L(\kpar)
\exp\bigl[-2\kappa(\kpar)d\bigr] t^{\sigma}_R(\kpar).
\label{factor}
\end{equation}
Here we replaced $|S_{rr^\prime}|^2$ by its explicit exponential
form with
\begin{equation}
\kappa(\kpar)=\left(\frac{2m\phi}{\hbar^2}+\kparrm^2\right)^{1/2},
\label{decay}
\end{equation}
where $\phi$ is the work function, and $d$ is the distance between
the reference planes assigned to the electrodes as shown in Fig.
\ref{reference}. All the information about the properties of
individual surfaces is described by the \emph{surface transmission
functions} (STF) $t^{\sigma}_L$, $t^{\sigma}_R$:
\begin{equation}\label{STF}
%t^{\sigma}_L(\kpar)=\sum_p\frac{\hbar\kappa}{mv_p}\left|\frac{B_r}{A_p}\right|^2,\;\;
%t^{\sigma}_R(\kpar)=\sum_q\frac{mv_q}{\hbar\kappa}\left|\frac{C_q}{B_{r^\prime}}\right|^2
t_L(\kpar)=\sum_p\left|\frac{B_r}{A_p}\right|^2,\;\;
t_R(\kpar)=\sum_q\left|\frac{C_q}{B_{r^\prime}}\right|^2,
\end{equation}
where the four amplitudes characterize the behavior of tunneling
eigenstates at the two surfaces (see Fig. \ref{reference}). The
different definitions of $t_L$ and $t_R$ are due to the fact that
they obey different boundary conditions. Specifically, $t_L$ and
$t_R$ are obtained by matching the Bloch wave functions with the
vacuum eigenstates, decaying and growing into the vacuum
respectively. The definition of $t_L$ implies the solution of a
scattering problem for the incoming wave with amplitude $A_p$, and
$B_r$ is the amplitude of this scattering eigenstate at the
reference plane in the vacuum. On the other hand, $t_R$ describes
an inverse scattering problem in which the exponentially decaying
wave in the vacuum with amplitude $B_{r^\prime}$ at the reference
plane is scattered on the right surface; here $C_q$ is the
amplitude of the outgoing Bloch state $q$ in the right electrode
for this ``eigenstate.'' Note that physically, this state is
forbidden because it grows to infinity in the vacuum, but it is
still a formal solution of the Schr\"odinger equation with the
specified boundary condition at infinity.

Each surface may be considered as forming the left or the right
interface of the tunnel junction. It is straightforward to show
that, just as in the case of a transmission matrix connecting the
propagating states in the two electrodes, \cite{Datta} the
requirement of flux conservation results in the reciprocity
condition $t_L(\kpar)=t_R(\kpar)\equiv t(\kpar)$ for any
(laterally periodic) surface, as long as the appropriate
normalization of the wave functions is chosen. Specifically, all
Bloch waves in the electrodes should be normalized for unit flux,
while the vacuum eigenstates should be normalized for unit
``imaginary flux'' $\kappa/m$. The easiest way to establish this
reciprocity condition is to prove it for free electrons in a
semi-infinite potential well, and then to use this free-electron
system as a second electrode in a junction. Since the total
transmission function for a two-electrode system is reciprocal,
\cite{Datta} this proves the above reciprocity condition for the
STF. Note that this proof does not rely on time reversal symmetry,
because time reversal also replaces $\kpar$ by $-\kpar$. However,
this symmetry gives another useful relation,
$t(\kpar,\mathbf{H})=t(-\kpar,-\mathbf{H})$ where $\mathbf{H}$ is
the external magnetic field.

It is easy to see from Eq. (\ref{STF}) with the chosen unit flux
normalization that the STF is proportional to the Fermi-level
value of the $\kpar$- and energy-resolved charge density for the
given spin, which is generated by the incoming Bloch states and
taken at the reference plane (any localized surface states are
excluded). Indeed, the energy-resolved charge density may be
written as
\begin{equation}\label{CD}
\rho_\sigma(\mathbf{r},E)=\frac{1}{N_\parallel}\sum_{\kpar}\rho_\sigma(\mathbf{r},\kpar,E)
\end{equation}
where the $\kpar$- and energy-resolved charge density is
\begin{equation}\label{kCD}
\rho_\sigma(\mathbf{r},\kpar,E)=\sum_p \left|\psi^F_{\sigma\kpar
p}(\mathbf{r})\right|^2
\end{equation}
where the summation is over the incoming Bloch states with the
given $\kpar$ and $E$, and the superscript $F$ emphasizes that the
scattering eigenstates $\psi^F_{\sigma\kpar p}$ are normalized so
that the incoming Bloch waves carry unit flux normal to the
interface. Setting $A_p=1$ in Eq. (\ref{STF}), we see that
$t(\kpar)$ is given by Eq. (\ref{kCD}) where $\mathbf{r}$ is taken
at the reference plane (since we neglect all components with
$\mathbf{G}_i\neq0$ at the reference plane,
$\rho(\mathbf{r},\kpar,E)$ does not depend on
$\mathbf{r}_\parallel$, the location within this plane).

Although we considered a vacuum tunneling barrier, the analysis
can be extended to other physically important cases of insulating
barriers.
%For the vacuum barrier, the main error
%originates from the (small) neglect of multiple scattering events,
%where a wave function crosses the barrier more than once. In the
%more general case, there is an additional requirement (always
%satisfied by a vacuum barrier)
Indeed, the main requirement for the validity of
Eqs.~(\ref{Smatrix},\ref{factor}) is that the complex band
structure be predominantly represented by a single evanescent wave
for each $\kpar$. At a minimum, this premise must hold for the
``active'' regions of the SBZ that contribute appreciably to the
tunneling current. This means that other tunneling states must
have a notably larger imaginary part of the wave vector compared
to the dominant one. In the case of a vacuum barrier, this amounts
to the neglect of all vacuum eigenstates with $\mathbf{G}_i\neq0$.
In practice this criterion is well satisfied for sufficiently
thick wide-gap tunnel barriers, such as $sp$-bonded oxides.
However, for any particular barrier, this assumption has to be
carefully verified by checking the complex band structure of the
barrier for the presence of additional slowly decaying states in
the active regions of the SBZ. It is important to note that the
area of the active region quickly shrinks as the barrier thickness
is increased,\cite{Mavro} indicating that the single evanescent
state criterion will often lead to an additional requirement of
``sufficient thickness'' of the barrier. Note that this
requirement is not related to a similar one stemming from the
neglect of back-scattering.

The factorization (\ref{factor}) allows us to study tunneling
between completely different electrodes. For practical purposes,
the STF can be calculated using $T_{\sigma}(\kpar)$ obtained for a
symmetric MTJ by factoring out the vacuum decay factor for the
given choice of reference planes, and then taking the square root.
The resulting STFs for different surfaces may then be convolved
with an appropriate vacuum decay factor, $\exp(-2\kappa d)$, to
obtain the transmission functions for asymmetric MTJs. In
particular, $T_{\sigma}(\kpar)$ for the antiparallel (AP)
magnetization of the two electrodes (and the TMR) can be found
from the up- and down-spin STFs calculated from the transmission
function for the parallel (P) magnetization.

\section{Generalization of Julli\`ere's formula}

Tunneling magnetoresistance is often discussed in terms of
Julli\`ere's formula\cite{Julliere}
\begin{equation}\label{Julliere}
    \mathrm{TMR}=\frac{2P_LP_R}{1-P_LP_R}
\end{equation}
where $P_L$, $P_R$ are the ``spin polarizations'' of the left and
right electrodes. Eq. (\ref{Julliere}) is derived from the
empirical expression $G^\sigma\propto\rho^\sigma_L\rho^\sigma_R$,
where $\rho^\sigma_L$, $\rho^\sigma_R$ are the ``tunneling
densities of states'' of the electrodes, and the spin polarization
is defined as
$P=(\rho^\uparrow-\rho^\downarrow)/(\rho^\uparrow+\rho^\downarrow)$.
The popularity of this formula is due to the fact that it usually
agrees reasonably well with experiment if the spin polarizations,
which are directly related to the ``tunneling densities of
states,'' are taken from Meservey-Tedrow experiments \cite{MT}
with the same barrier. However, the validity of Julli\`ere's
formula has been debated for a long time, and the reasons for its
apparent agreement with experiment are unclear. The physical
meaning of the ``tunneling density of states'' is also unclear,
but it is obvious both from elementary quantum mechanics and from
experiments that the tunneling properties of a magnetic
heterostructure are determined not by the ferromagnet alone, but
rather by the ferromagnet/barrier combination and by the structure
of the interface. A number of explicit first-principles
calculations for idealized MTJs without disorder (see, e.g., Ref.
\onlinecite{Maclaren97}) confirmed this fact. However, it was
suggested \cite{MathUm} that phase decoherence due to disorder
which is always present in realistic MTJs may recover the
factorization of the tunneling conductance in a product of
``transport densities of states,'' which are essentially equal to
the regular densities of states at the surfaces of the electrodes
if there are no resonant localized states in the barrier.
Moreover, it was shown \cite{TP98} within a single-band tight
binding model that in the limit of strong disorder one recovers
Julli\`ere's formula (\ref{Julliere}) by identifying $P_L$, $P_R$
with the measurable spin polarizations of the tunneling current
for the same electrode/barrier systems. Therefore, it seems that
there are good reasons for the widespread use of Julli\`ere's
formula, and it is highly desirable to elucidate these reasons.

Let us explore the connection between Eq. (\ref{factor}) and
Julli\`ere's formula (\ref{Julliere}). In Eq. (\ref{factor}) the
simple product of the ``tunneling densities of states'' is
replaced by a convolution of STFs, which explicitly include the
effects of bulk densities of states and of the surface structure.
Thus we can consider Eq. (\ref{factor}) as a generalization of
Julli\`ere's formula for an ideal MTJ with no disorder.
%We shall show below that Eq.~(\ref{factor}) in many
%circumstances is nearly exact.

However, we may go further and identify limiting cases where Eq.
(\ref{factor}) can be directly related to Julli\`ere's formula,
providing formal definitions of the ``tunneling densities of
states'' of the electrodes appropriate for these cases. First,
consider the case of a disordered insulating barrier. Such a
barrier may be characterized by its eigenstates, half of which are
decaying from left to right, and the other half from right to
left. Although these eigenstates do not have a conserved $\kpar$
any more, it is still clear that tunneling will be dominated by
Feynman paths that do not ``loop back,'' because each path carries
a weight decaying exponentially with its length (see below).
Therefore, we may still write an expression similar to Eq.
(\ref{Smatrix}) neglecting back-scattering, but now we should sum
up over all barrier eigenstates (now defined in real space).
Within this formulation $S_{rr^\prime}$ is still diagonal because
it describes the decay of a single eigenstate.

The weight of a Feynman path in the imaginary-time functional
integral often used for tunneling problems\cite{Coleman} (with
Euclidean action written in its reduced Maupertuis form; see,
e.g., Ref. \onlinecite{Katsnelson}) is given by
$\exp\left[-\int\kappa(l)dl\right]$ up to a prefactor, where the
integral is taken along the path, $\kappa=[2m(V-E)]^{1/2}$ and
$V(\mathbf{r})$ is the potential. In an ordered insulator many
paths with similar weights contribute to the path integral
resulting in the formation of the complex band structure. However,
in a disordered insulator the tunneling current may be dominated
by Feynman paths running close to a relatively small number of
``easy'' paths with locally maximum weights, i.e. by
imaginary-time classical paths.\cite{Coleman,Katsnelson} If there
is only one such channel or one class of channels with similar
properties (e.g., due to surface roughness), Eq. (\ref{factor})
will produce Julli\`ere's formula where $\rho^\sigma$ is simply
the Fermi-level value of the energy-resolved charge density given
by Eq. (\ref{kCD}) integrated over $\kpar$ and taken at some
reference point within the channel. (Now each term describes the
scattering eigenstate corresponding to the single incoming Bloch
wave with the given $\kpar$.) Like STF, this quantity does not
depend on the properties of the other side of the barrier. This
conclusion agrees with the results of Ref. \onlinecite{TP98}
showing that the tunneling current through a strongly disordered
barrier is dominated by a small number of random configurations,
and that Julli\`ere's formula is also recovered in this limit.

Now consider the case when disorder is weak close to the
interfaces, but remains strong in the insulator. Obviously, the
S-matrix of the disordered insulator in $\kpar$ representation
will be essentially a random matrix, and after averaging Eq.
(\ref{factor}) thus yields Julli\`ere's formula with
$\rho^\sigma\propto\sum_{\kpar}t^\sigma(\kpar)$. This case is the
easiest from the computational point of view, because the STFs may
be directly calculated for a $\kpar$-conserving MTJ.

It is instructive to compare this result with the conclusions of
Mathon and Umerski \cite{MathUm} on the applicability of
Julli\`ere's formula obtained using the transfer Hamiltonian
formalism. Our approach shares in common with Ref.
\onlinecite{MathUm} the neglect of multiple reflections across the
junction. However, the assumption of constant matrix elements
(hopping integrals) for all Bloch waves made in Ref.
\onlinecite{MathUm} completely removes all physical effects
connected with orbital- and spin-dependent bonding at interfaces.
This obviously contradicts the experimental findings showing that
the spin polarization of the tunneling current and
magnetoresistance strongly depend on the type of barrier
used.\cite{TML} In our approach, the STFs for the electrodes allow
us to encapsulate the effects of the interface structure and
provide the proper dependence of the tunneling current on barrier
type. Julli\`ere's formula obtained in the limiting case of full
decoherence inside the insulator is expressed in terms of the spin
polarization actually measured in the Meservey-Tedrow experiment
(assuming that the superconductor acts as an ideal, non-biased
spin detector).

Finally, for very thick $\kpar$-conserving barriers the tunneling
current may be carried predominantly by a close vicinity of some
special $\kpar$-points in the SBZ (e.g., the $\Gamma$ point). In
this case, the tunneling density of states is simply equal to the
value of $T(\kpar)$ at this $\kpar$.

It is not clear \emph{a priori} whether any one of these three
limiting cases is directly applicable to realistic MTJs, although
it seems that disorder in the insulator together with the
``channelization'' of the tunneling current are both likely to
play a major role. However, the emergence of Julli\`ere's formula
in these different scenarios suggests that it may actually have a
rather wide range of applicability. In general, the ``tunneling
density of states'' should be identified with some appropriately
averaged energy-resolved charge density taken at the Fermi level
at a sufficient distance from the interface within the barrier.
Unlike the bulk density of states, this function fully takes into
account the relevant properties of the surface.

\section{Tunneling from clean and oxidized C\lowercase{o} (111) surfaces through vacuum into A\lowercase{l}}

We calculated the transmission functions using the principal-layer
Green's function approach \cite{Turek-book} based on the
tight-binding linear muffin-tin orbital method (TB-LMTO) in the
atomic sphere approximation (ASA) and the transmission matrix
formulation of Ref. \onlinecite{Kudr}. All the atomic potentials
were determined self-consistently within the supercell approach
using the TB-LMTO-ASA method. The vacuum barrier was modeled using
empty spheres in the positions corresponding to the continuation
of the crystal lattice of the electrodes. We have also performed
full-potential LMTO calculations \cite{FLMTO} which confirmed all
main features of the band structure of the oxidized Co (111)
surface discussed below.

We checked the validity of factorization (\ref{factor}) by
calculating $T_{\sigma}(\kpar)$ for (100) and (111)-oriented fcc
Co electrodes with parallel magnetizations, taking the square
root, and convolving $t_\uparrow(\kpar)$ with
$t_\downarrow(\kpar)$. Then, the result was compared with the
independent calculation for the antiparallel configuration in a
range of energies. The agreement was always excellent (better than
1\%), except for a couple of specific energies for a (100) MTJ
with four vacuum ``monolayers'' where narrow resonances appear in
the minority channel. \cite{Wunnicke} If the vacuum barrier is
extended to 8 monolayers, excellent agreement is restored.
%This means that the perturbation
%theory approach breaks down in the presence of resonance interface states, as expected.

Using the factorization (\ref{factor}) we investigated the SP of
the conductance from ferromagnetic electrodes to a non-magnetic
material, Al (111), which served as a detector of the tunneling SP
in the spirit of the Meservey-Tedrow experiments. \cite{MT} As
expected, the calculated STF of Al is free-electron-like, having
almost perfect Gaussian shape originating from the vacuum decay
factor up to the reference plane. Therefore, this surface may be
considered as equally transparent for all Bloch waves, and the
total transmission function for a MTJ with Al spin-detector
electrode is essentially a product of the other electrode's STF
and the vacuum decay factor.

\begin{figure}
\begin{center}
\epsfig{file=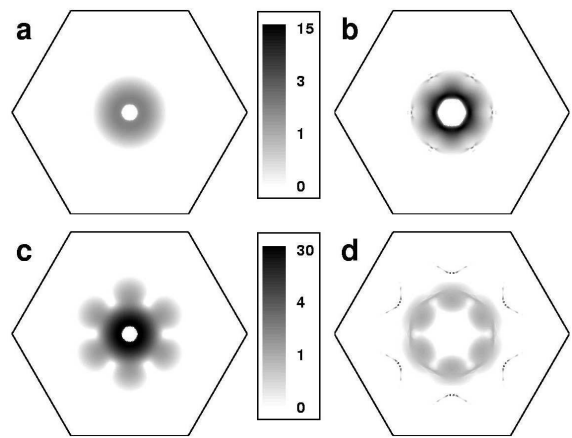,width=.42\textwidth}
\end{center}
\caption{$\kpar$-resolved transmission (logarithmic scale) from
clean and oxidized (111) Co surfaces through vacuum into Al. (a)
Clean surface, majority spin. (b) Clean surface, minority spin.
(c) Oxidized surface, majority spin. (d) Oxidized surface,
minority spin. The vacuum layer thickness is 2~nm for clean and
1.7~nm for oxidized Co surface. The first surface Brillouin zone
is shown. Units are $10^{-11}$ for (a), (b) and $10^{-14}$ for
(c), (d).} \label{transmission}
\end{figure}

\begin{figure}
\begin{center}
\epsfig{file=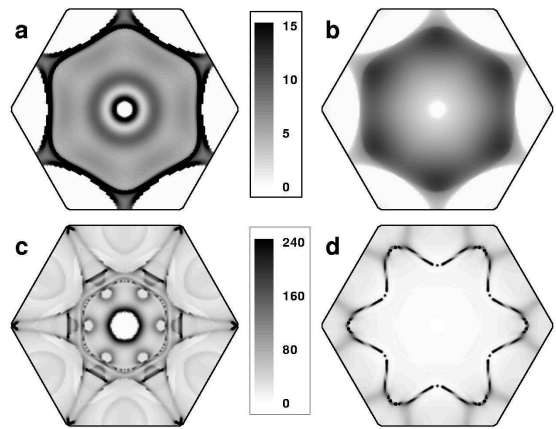,width=.41\textwidth,clip=true}
\end{center}
\caption{$\kpar$-resolved local DOS at $E_F$ (arbitrary units) for the oxidized Co surface:
(a) S6 layer, majority spin;
(b) S1 layer, majority spin;
(c) S6 layer, minority spin;
(d) S1 layer, minority spin.}
\label{partialDOS}
\end{figure}

First, we discuss the properties of a Co/vacuum/Al MTJ with a
clean Co (111) surface. Figs.~\ref{transmission}a,b show the
$\kpar$-resolved transmission for the majority- and minority-spin
electrons within the SBZ of Co (111). The Fermi surface of Co
viewed along the [111] direction has holes close to the
$\bar\Gamma$ point with no bulk states in both spin channels,
which results in zero conductance in this area. The majority-spin
transmission (Fig.~\ref{transmission}a) varies rather smoothly and
is appreciable over a relatively large area of the SBZ. On the
other hand, the minority-spin transmission
(Fig.~\ref{transmission}b) has a narrow crown-shaped ``hot ring''
around the edge of the Fermi surface hole. The analysis of layer
and $\kpar$-resolved density of states (DOS) shows that it is not
associated with surface states, \cite{Wunnicke} but rather with an
enhancement of bulk $\kpar$-resolved DOS near the Fermi surface
edge (compare Fig.~\ref{transmission}b with
Fig.~\ref{partialDOS}c).

As seen from Figs.~\ref{transmission}a,b, the Fermi surface hole
is smaller for majority spins. Therefore, the conductance should
become fully majority-spin polarized in the limit of very thick
barriers. However, since the Fermi surface hole is also quite
narrow for minority spins, positive SP is only achieved at very
large barrier thicknesses $d\sim10$~nm, while for typical values
of $d\sim2$~nm the SP is about $-60\%$ and depends weakly on $d$.

The oxidized Co surface was modeled by an O monolayer placed on
top of the Co(111) electrode. The equilibrium atomic structure of
this surface was found using the pseudopotential plane-wave method
\cite{CASTEP} within the generalized gradient approximation. We
used both types of stacking: ABCA and ABCB, where the last symbol
designates the position of the O monolayer. The O atoms are
assumed to lie in symmetric positions above the second (S2) or
first (S1) sub-surface Co layer, respectively (these layers are
shown in Fig. \ref{reference} in the absence of the O monolayer).
We allowed the O layer and two Co layers (S1 and S2) to relax in
the direction normal to the surface, while the positions of atoms
in deeper layers (S3,\dots) were kept fixed. The energies of both
equilibrium configurations of oxygen monolayers were found to be
very close to each other. All results of interest in the present
context are quite similar for these two stackings. Below all
specific data are given for the ABCB stacking. The equilibrium
interlayer distances were found to be 2.14~\AA~between layers S3
and S2, 2.18~\AA~between S2 and S1, and 1.08~\AA~between S1 and O
layers, compared to 2.07~\AA~between the adjacent Co layers in the
bulk. The Co-O bond length is equal to 1.82~\AA.

Presence of oxygen at the surface of cobalt raises the question of
whether electron correlations similar to those characteristic for
transition-metal oxides may be strong enough to induce significant
changes in the band structure at the oxidized Co surface. However,
the enhancement of correlations in oxides is due to much weaker
screening of Coulomb interaction compared to the metallic state.
On the other hand, cobalt atoms below the oxygen monolayer
preserve the close-packed configuration of bulk cobalt except for
the top three nearest neighbors being absent out of twelve.
Therefore, it is reasonable to expect that screening of Coulomb
interaction in the $3d$ shell is not much weaker compared to the
bulk. For this reason, we believe that LDA electronic structure of
the oxidized Co surface is correct as far as the main features are
concerned.

\begin{figure}
\begin{center}
\epsfig{file=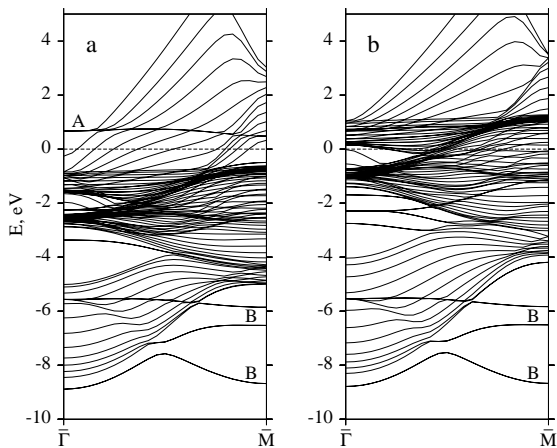,width=.41\textwidth}
\end{center}
\caption{Band dispersions along the $\bar\Gamma\bar M$ direction
for (a) majority and (b) minority spin electrons.
Energy is referenced from $E_F$.
The bonding Co-O surface bands are marked B, and the
pure antibonding surface band is marked A.}
\label{bandplots}
\end{figure}

The oxygen monolayer dramatically changes the electronic structure
of the underlying Co layer making this layer almost
magnetically-dead. This change can be understood from band
dispersion plots shown in Fig.~\ref{bandplots}. For each spin, the
free-standing oxygen monolayer has three energy bands deriving
from $2p$ states, each doubly degenerate due to $\sigma_z$
reflection symmetry ($z$ is the axis normal to the surface). When
the monolayer is deposited onto the Co surface, the degeneracy is
lifted, and two sets of three bands each are formed corresponding
to bonding and antibonding mixing of oxygen and cobalt orbitals.
The three bonding bands (marked B in Fig.~\ref{bandplots}) lie
well below the bulk Co $3d$ band, whereas the antibonding states
are close to the Fermi energy $E_F$. As a result of this bonding
the local DOS for the S1 layer at $E_F$ is strongly reduced, so
that, according to the Stoner criterion, magnetism in this layer
is almost completely suppressed. The magnetic moment of Co atoms
in the S1 layer is only 0.17 $\mu_B$.

%If the resulting surface states remained pure (did not mix with bulk states), the
%$\kpar$-resolved DOS for surface layers would only show infinitesimally thin curves
%corresponding to the intersection of surface bands with the Fermi level. This situation
%occurs if no bulk states are available for the given $\kpar$. However, even
%if the bulk states do couple to surface states, the coupling may be weak (owing
%to spatial separation or to some selection rules enforced by symmetry).

%Fig.~\ref{bandplots} shows that some antibonding bands readily mix with the bulk states,
%and some do not. This is due to a selection rule which is exact along the $\bar\Gamma\bar M$ line.
%The present case of oxidized Co surface presents a good example of such selection rules.
%This line is the projection of a reflection plane. Therefore, all states with
%$\kpar$ on this line are either even or odd with respect to this reflection.
%Although the states with general $\kpar$ do not obey strict selection rules,
%this behavior is largely preserved throughout the entire SBZ.
%, since $\bar M$ points form a hexagonal star.
%Below we denote the bands as ``even'' or ``odd'' according to their
%symmetry with respect to $\bar\Gamma\bar M$ reflection, but one has to keep in mind that the
%selection rule for other directions is not exact.

Transmission of propagating bulk states from the electrode through
the barrier is very sensitive to the degree of mixing of these
states with the antibonding surface states. This mixing is
controlled by a selection rule which follows from the fact that
all bands can be classified as ``even'' or ``odd'' according to
their symmetry with respect to $\bar\Gamma\bar M$ reflection
($\bar\Gamma\bar M$ is the projection of a reflection plane).
Although this classification is exact only along the
$\bar\Gamma\bar M$ direction, it is approximately valid throughout
the entire SBZ.

According to this classification, two of the three surface bands
are even, and one is odd. On the other hand, the
free-electron-like band of bulk Co which forms the only Fermi
surface sheet for majority-spin electrons is even, while the
minority-spin states on the Fermi surface sheet closest to the
$\bar\Gamma$ point are odd. Even and odd bands are orthogonal and
can not mix.

This selection rule results in the principal difference between
the majority- and minority-spin transmission. The majority-spin
bands are shown in Fig.~\ref{bandplots}a. One even and one odd
antibonding surface bands (marked A) are degenerate at the
$\bar\Gamma$ point (at about 0.8~eV above $E_F$). At a short
distance from the $\bar\Gamma$ point both bands enter the
continuum of bulk states. The odd band does not mix with the bulk
states along the $\bar\Gamma\bar M$ line and remains almost flat
due to repulsion from a lower-lying band.
%In going from $\bar\Gamma$ to $\bar K$ the oxygen content of these states
%gradually changes from $p_{xy}$ to $p_z$.
On the other hand, the even band readily mixes with the
free-electron-like majority-spin band crossing the Fermi level and
completely loses its surface localization.
%We see in Figs.~2a and \ref{partialDOS}b that the bulk free-electron-like band
%has a significant amount of Co(S1) character at the Fermi level,
%although it is somewhat suppressed close to the $\bar\Gamma$ point.
%Naturally, Co(S1) character is accompanied by O character.
This is evident from Figs.~\ref{partialDOS}a,b which show the $\kpar$-resolved DOS of majority-spin
electrons for the bulk-like S6 and surface S1 Co layers.
The DOS for the S1 layer (Fig.~\ref{partialDOS}b) is appreciable in the entire area of the SBZ where bulk states
are available and has no sharp features that might indicate localized surface states. This implies
that the bulk majority-spin states extend to the very surface of the electrode and therefore can
readily tunnel through the barrier.
%The partial oxygen-layer majority-spin $\kpar$-resolved DOS (not shown) is nearly
%homogeneous over the entire area of the SBZ where bulk states are available.
%Finally, the third surface band falls completely into the bulk $3d$ band and does not approach the Fermi level.

The situation is very different for minority-spin states.
Although the odd surface band is again almost flat and lies above $E_F$,
the even surface band crossing the Fermi level does not mix with the odd minority-spin
band. As a consequence, the
$\kpar$-resolved DOS for Co(S1) layer (Fig.~\ref{partialDOS}d) is large only along the curve lying
at the periphery of the SBZ where the Co-O antibonding surface band crosses the Fermi level (oxygen DOS looks very similar).
%There is also a region of appreciable DOS along the edge of the SBZ where the surface bands mix
%with bulk states in other sheets of the Fermi surface. Transmission through all these states
%is suppressed due to their fast decay in the vacuum.
As a result, the bulk minority-spin states responsible for most tunneling transmission from the clean
surface only extend up to the S2 layer, encountering a band gap in the S1 and oxygen layers.
%which is equivalent to adding an additional tunneling barrier for minority-spin states.
%On the other hand, the transmission through the
%surface resonances is suppressed because they are very far from the $\bar\Gamma$ point.
%The total spin-polarization of the tunneling current from the oxidized surface is thus almost 100\% positive.
%Thus, oxidation of the Co (111) surface creates a spin-filter effect due to purely band-structure
%effects, demonstrating the crucial role of surface atomic and electronic structure in SDT.
%Thus, the result of the Co-O bonding-antibonding splitting and poor mixing of odd bulk minority-spin states
%with even surface states is almost complete removal from the Fermi energy (in the S1 and O layers) of those
%minority-spin states that dominate in the STF of the clean surface (compare Figs.~\ref{partialDOS}d and
%\ref{partialDOS}c). This is equivalent to adding
%a layer of insulator for minority-spin states. On the contrary, although the majority-spin states
%are pushed a little farther from the $\bar\Gamma$ point, they are not removed from the Fermi energy
%(compare Figs.~\ref{partialDOS}b and \ref{partialDOS}a),
%because the relevant surface band mixes strongly with the majority-spin states.
Thus, an additional
tunneling barrier is introduced in the minority-spin channel, and the total SP of the tunneling
transmission becomes
almost 100\% positive, which is evident from Figs.~\ref{transmission}c,d.

The predicted effect of interface bonding is not limited only to
the Co(111) surface. We have also calculated the transmission from
clean and oxidized Co(100) and Ni(111) electrodes and found that
surface oxidation reverses the SP due to the bonding between Co or
Ni with O. As it was shown earlier, the reversal of the SP also
occurs for the Fe(100) surface. \cite{TOP}

\section{Conclusion}

We have shown that the problem of calculating of the transmission
function for a sufficiently thick insulating barrier is reduced to
the solution of three separate problems, namely the penetration of
the bulk wave functions into the barrier from both sides, and the
behavior of the evanescent barrier eigenstates. This separation
provides a natural generalization of Julli\`ere's formula. We
identified three limiting cases when the original Julli\`ere's
formula is recovered. The ``tunneling density of states'' in this
formula is identified with an appropriately averaged
energy-resolved charge density generated by the bulk Bloch states
within the barrier and taken at the Fermi level.

Using the factorization of the transmission function into a
product of surface transmission functions and a barrier decay
factor we calculated the spin polarization of the tunneling
current from clean and oxidized Co (111) surfaces through vacuum
into Al. We showed that the bonding between Co and O atoms at the
oxidized surface controls SDT by creating an additional barrier
for minority-spin electrons, which results in a reversal of the
spin polarization.

Experimentally, the reversal of the SP associated with surface
oxidation may be detected using spin-polarized STM measurements.
\cite{SP-STM} Since the ferromagnetic tip is sensitive to the SP
of the total local DOS above the surface (see, e.g., Ref.
\onlinecite{Heinze}), the TMR in the system surface/vacuum/tip
should change sign when the Co surface is oxidized. In other
words, for the clean Co (111) surface the tunneling current should
be higher when the magnetizations of the tip and the surface are
aligned parallel (the dominating minority channel is then open),
but for the oxidized surface it should be higher for the
antiparallel configuration.

\begin{acknowledgments}

We are grateful to V. P. Antropov for useful discussions. This
work is supported by NSF (DMR-0203359 and MRSEC DMR-0213808) and
the Nebraska Research Initiative. MvS was supported by DARPA and
ONR. Work at Sandia National Laboratories is supported by
%the Office of Basic Energy Sciences, Division of Materials Sciences,
U.S. Department of Energy under contract No. DE-AC0494AL85000.

\end{acknowledgments}

\end{document}